# Full Momentum and Energy Resolved Spectral Function of a 2D Electronic System

Joonho Jang[1], Heun Mo Yoo[1], Loren Pfeiffer[2], Ken West[2], K.W. Baldwin[2], and Raymond Ashoori[1]

[1]*Department of Physics, Massachusetts Institute of Technology, Cambridge, MA 02139, USA*
[2]*Department of Electrical Engineering, Princeton University, Princeton, NJ 08544, USA*

**The single-particle spectral function measures the density of electronic states (DOS) in a material as a function of both momentum and energy, providing central insights into phenomena such as superconductivity and Mott insulators. While scanning tunneling microscopy (STM) and other tunneling methods have provided partial spectral information, until now only angle-resolved photoemission spectroscopy (ARPES) has permitted a comprehensive determination of the spectral function of materials in both momentum and energy. However, ARPES operates only on electronic systems at the material surface and cannot work in the presence of applied magnetic fields. Here, we demonstrate a new method for determining the full momentum and energy resolved electronic spectral function of a two-dimensional (2D) electronic system embedded in a semiconductor. In contrast with ARPES, the technique remains operational in the presence of large externally applied magnetic fields and functions for electronic systems with zero electrical conductivity or with zero electron density. It provides a direct high-resolution and high-fidelity probe of the dispersion and dynamics of the interacting 2D electron system. By ensuring the system of interest remains under equilibrium conditions, we uncover delicate signatures of many-body effects involving electron-phonon interactions, plasmons, polarons, and a novel phonon analog of the vacuum Rabi splitting in atomic systems.**

Inside of semiconductors and metals, negatively-charged electrons interact with each other and with positively-charged ions and impurities, creating complex many-particle systems (*1*). Remarkably, despite the strong electron-electron Coulomb repulsions inside of solids, Landau-Fermi liquid theory predicts that in a degenerate Fermi system, electrons near Fermi level $E_F$ behave as weakly interacting quasiparticles (*1*, *2*). These quasiparticles have a modified energy and mass compared with non-interacting electrons and have finite lifetimes for losing their identities as a well-defined particles. The success of the theory stems from the fact that the degenerate electrons below the Fermi level screen the Coulomb potential between particles and restrict the phase space available for excited electrons to scatter. In cases for which the interactions of electrons become dominant, such as for systems of low electron density and/or under strong magnetic fields, the description in terms of weakly interacting quasiparticles of the Fermi liquid theory no longer accurately describes the system; electrons start to show strongly correlated behaviors, accompanied by unscreened, effectively enhanced electron-phonon (e-ph) and electron-electron (e-e) interactions. This phenomenon has a significant influence on many experimentally important quantities, e.g. transport, thermodynamic, and optical properties, in metals and insulators, and may lead to exotic phases such as superconductivity and Fractional Quantum Hall effect (*3*, *4*).

The spectral function, the distribution of single-particle electronic states in energy and momentum space, is a fundamental physical quantity that directly reflects underlying many-body interactions. Physicists have extensively developed the theory of the spectral function of electron liquid and strongly correlated electron systems (*1*, *2*, *5*, *6*). Established tools such as STM and ARPES have contributed greatly to understanding of electronic structure of solids by measuring the spectral function or quantities related to it (*7*, *8*). However, these methods only probe the electronic systems at the sample surface and do not work if the sample is electrically insulating. There has thus been no means for accurate determination of the spectral function of high-mobility systems created by molecular beam epitaxy, a long-time testbed for understanding fundamental physics and device applications due to their high quality and controllability. Most of the experimental studies in these systems utilize transport or thermodynamic probes that generally only have sensitivity to the average thermodynamic behavior of electrons near the Fermi energy.

In this work, we demonstrate a unique method of tunneling spectroscopy, momentum and energy resolved tunneling spectroscopy (MERTS), that utilizes a fine-control of the momentum and energy of tunnel-injected electrons and directly measures the spectral functions of 2D electron system (2DES) in a quantum well (QW) at various densities including near-depleted and fully-depleted (insulating) regimes. This technique dramatically visualizes the pronounced e-ph and e-e interaction effects deep inside solids. MERTS has comparable energy and momentum resolution to ARPES systems (*7*, *9*), a capability that is critical for investigation of the subtle e-e interaction effects near the Fermi level. Importantly, MERTS is capable of measuring E-k spectra in extreme conditions, where the deviation from the simple Fermi-liquid picture becomes important: **(i)** in insulating regimes near depletion or even in full depletion (**Fig. 2C**), and/or **(ii)** under strong perpendicular magnetic fields (**Fig. 3**).

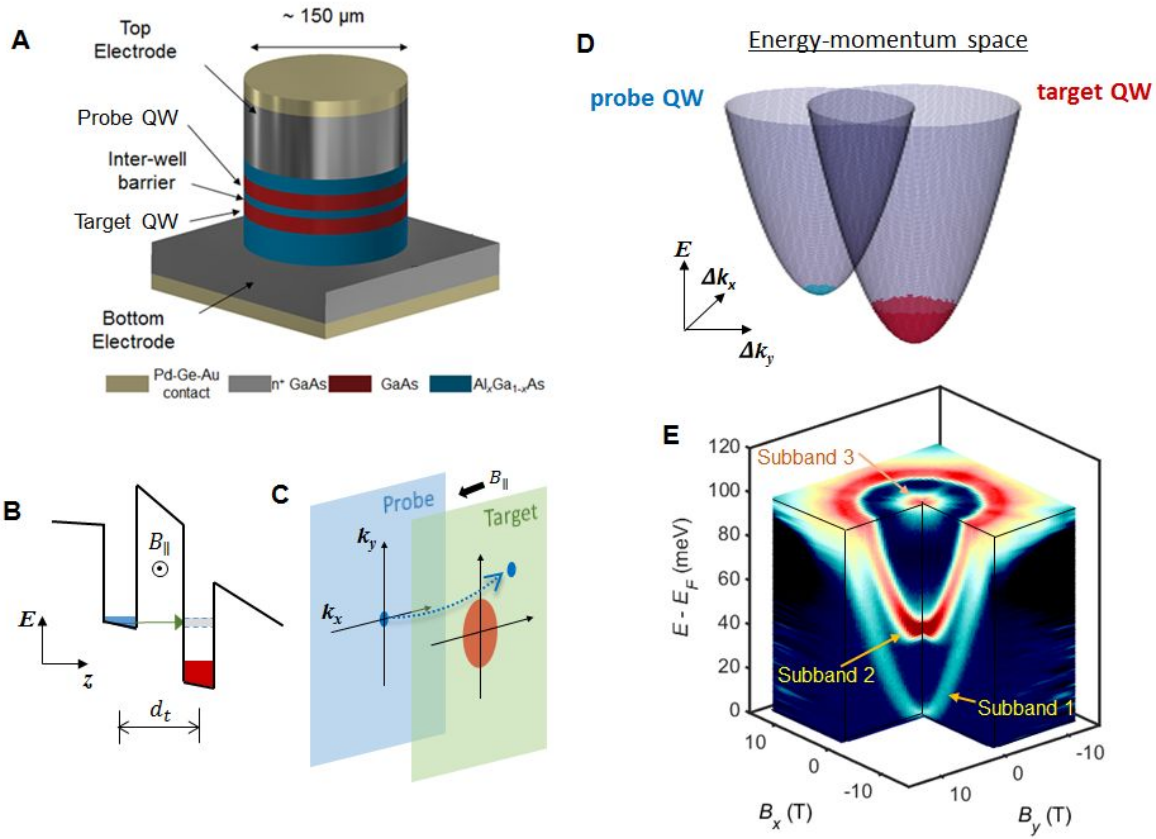

**Figure 1 | Schematics of tunneling device and principles of energy-momentum selection process.** **(A)** The vertical tunneling device used in the experiment. Two GaAs QWs (width of 20 nm each) are separated by a $Al_{0.8}Ga_{0.2}As$ potential barrier (6 nm). Electrons in the top (probe) QW with nearly zero planar momentum probe electronic states in the bottom (target) QW. **(B)** The injection of an electron packet probes only empty available states in the target layer. **(C)** Diagram for explaining the momentum selection mechanism. The in-plane field generates a momentum boost in the tunneling process to displace the zero planar momentum into $k_{final} = eB_{\parallel}d_t/\hbar$ in the bottom QW. **(D)** E-k dispersions in the presence of the in-plane field. In (B-D), the occupied states are shown in blue and red colors for probe and target layers, respectively. **(E)** Measured spectra with $B_{\parallel}$ along the crystallographic axis of [100] and [010] of GaAs. Multiple unoccupied QW subbands are visible.

In MERTS, we controllably generate a collimated packet of electrons with precisely-defined energy and momentum from an adjacent probe layer (a 2D QW) and, by means of tunneling, inject the electrons into the unoccupied quantum states of the target 2D system to measure the target 2D system's spectral function. Electrons from this collimated beam can tunnel into the target 2D electron system only when unoccupied states are available with the same energy and momentum, resulting in *E-k* selectivity of the tunneling process (see **Fig. 1**). The tunneling probability therefore represents the spectral function for adding a particle into the 2D system (*1, 2*). For high *E-k* resolution, we prepare electrons with a very small Fermi surface in the probe layer (containing a very low density of electrons $n \sim 3 \times 10^{10} cm^{-2}$ with $E_F \sim 1.0\ meV$ and $k_F \sim 0.004\ A^{-1}$; these numbers set the resolution of the technique) at E=0 and k=0 ($\Gamma$-point in the GaAs conduction band minimum), and translate the momentum of the packet using a magnetic field applied $B_\parallel$ perpendicular to the tunneling direction giving electrons a momentum shift of $\Delta k_{x,y} = eB_{\parallel y,x} d_t/\hbar$ in the tunneling process (*10, 11*). Here, $\Delta k_{x,y}$ is the momentum translation of the tunneling electrons and $d_t$ is the tunneling distance (**Fig. 1B-C**). This effect can be understood semi-classically as the momentum gain of an electron in traversing the tunnel barrier due to the magnetic field induced Lorentz force. In previous studies (*11, 12*) involving large Fermi surfaces in both the probe and target wells, the tunneling measurements involved complicated convolutions of two systems and electrical conductivity through both layers, making it unfeasible to obtain direct spectral information. In contrast, in MERTS we achieve full momentum and energy resolution by adapting a pulsed tunneling method (*13*) that requires no direct electrical contact to the system of 2 quantum wells and does not rely on conductivity of the 2D systems; the method functions even if either or both of the 2D systems are electrically insulating. The method has important elements that contributed to the success of MERTS by allowing us to **(i)** maintain the small Fermi surface of source electrons (probe layer) into nearly a single point and **(ii)** to eliminate lateral transport in either layer to ensure uniform tunneling at all points in the plane.

An example of tunneling spectra of a 2D electron system in a GaAs/AlGaAs QW (**Fig. 1A**) measured with $B_\parallel$ applied along [100] (x-axis) and [010] (y-axis) is plotted in **Fig. 1E**. We have verified the high fidelity of the momentum and energy conservation in the tunneling process by comparing to a numerical simulation in same experimental parameters (see **Fig. S3**). The MBE grown wafer is specifically designed so that the probe well has the low density described above. We vary the density of the target 2D system by applying a DC voltage $V_b$ across the top and bottom electrodes. An additional voltage pulse with a precisely controlled amplitude induces the tunneling voltage $V_t = E_t/e = (E - E_F)/e$ between the source and target layer (*13*), followed by a pulse with opposite polarity that retrieves the tunneled electrons to bring the entire system back to equilibrium for next tunneling pulses (see Materials and Methods). The tunneling current is determined by measuring the displacement charge induced by the voltage pulses using capacitively coupled HEMT amplifiers. In MERTS, the electronic DOS is directly proportional to the tunneling current *I*, in contrast to conventional tunneling measurements using a 3D electrode in which the DOS is proportional to $dI/dV$.

In **Fig. 2A-C**, we display the *E-k* distributions (dispersions) of electronic states at various carrier densities $n$. We calibrate the momentum axis, proportional to $B_\parallel$, based on knowledge of the tunneling distance $d_t$ determined from numerical simulations of the wavefunctions in each quantum well (designed to be $d_t = 26\ nm$, but changing slightly with $V_b$). Adjusting $V_b$ to more negative values decreases the electron density of the target QW and eventually fully depletes it. The spectrum labeled as an empty well in **Fig. 2C** clearly shows the depletion of electrons from the conduction band, visualized by the bottom of the conduction band edge rising above the Fermi level. The absence of occupied extended electron states in the QW is a clear signature of a system entering an insulating phase, consistent with our capacitance measurements of the structure.

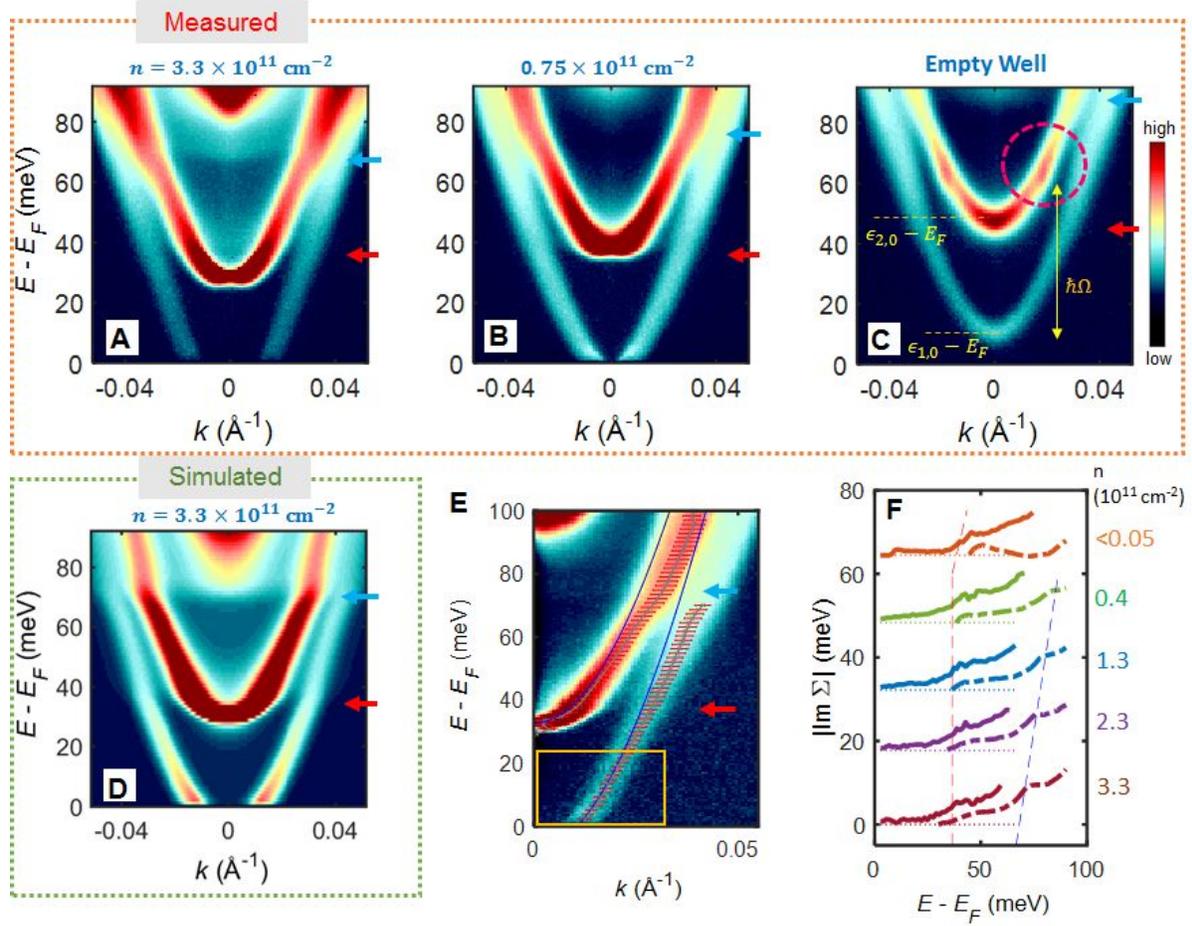

**Figure 2 | Measured MERTS Spectra at various densities** $n$ **and extraction of self-energy.** **(A-C)** Tunneling spectra measured at 1.5 K for various target QW densities (see Fig. S7 for a full data set). In (A) three subbands are visible due to finite width of the target well. Frame (C) shows data for a gate voltage at which the target well first fully depletes. **(D)** A tunneling spectrum simulation with a self-energy term that accounts for impurity broadening and electron-phonon interactions. The red arrows are placed a LO phonon energy (36 meV) above the Fermi energy in (A), (B) and (D), and 36 meV above the bottom of the 1st subband in (C). The blue arrows in each figure are 36 meV above the 2nd subband. **(E)** A measured *E-k* spectrum at $n = 2.3 \times 10^{11}\ cm^{-2}$. The blue curves displays the bare band dispersion $\varepsilon_k - E_F$ expected from detailed band structure calculations (*14*). We extract the self-energy by measuring the peak locations $k_{peak}$ and half-widths $\Delta k$ of MDCs (curves cut along the momentum axis at energy $E$). The red bars are centered at $k_{peak}$ and have widths of $\Delta k$. We extract the self-energy using $Re\ \Sigma = E - \varepsilon(k_{peak})$ and $|Im\ \Sigma| \approx (\Delta k - 0.0032\ A^{-1}) \times \hbar v_{g0}$ (*15–17*), where the subtraction of 0.0032 is to take account of the finite size of FS of the probe QW, and $v_{g0}$ is the bare group-velocity (see section 5 of Supplementary Materials). The yellow box marks an area of the spectrum considered in Fig. 4. **(F)** Extracted $Im\ \Sigma$ s at various densities. The solid curves are for data from the 1st subband, and the dashed curves are for 2nd subband. The red and blue dashed (nearly vertical) lines are guides-to-the-eyes for the abrupt increases of $|Im\ \Sigma|$ due to e-phonon interactions. These are the same with energy locations pointed by the red and blue arrows in (A), (D) and (E), respectively. Curves are offset for clarity, and the thin horizontal dotted line indicates zero for each curve.

The spectra also display quantized subbands due to the confinement potential in the growth direction (z-axis). We identify the three lowest subbands in the spectrum; noticeably, the intensity of the spectrum grows stronger for subbands lying higher in energy because the higher subband wave functions penetrate deeper into the barrier and thus have stronger tunnel coupling to the probe well. In a model of the bands that excludes e-e and e-ph interactions, the single particle energies $\varepsilon_{n,k}$ follow $\varepsilon_{n,k} = \varepsilon_{n,0} + \hbar^2/2m_{eff} \times (k_x^2 + k_y^2) \times (1 + \eta|k|^2)$, shown by the blue curves drawn in **Fig. 2E**. Here, $\varepsilon_{n,0}$ is the confinement energy of n-th subband, $k_x$ and $k_y$ are planar momenta, and $m_{eff}$ is the band mass, and $\eta$ (~ -43.7 $Å^2$) is a dimensionless parameter that accounts for band non-parabolicity (*14*). However, a more comprehensive picture beyond this effective mass approximation comes from the description of interacting electrons with the spectral function (*1, 2, 7, 18*):

$$A(E,k) = \frac{1}{\pi} \frac{|Im\ \Sigma|}{(E - \varepsilon_{n,k} - Re\ \Sigma)^2 + (Im\ \Sigma)^2} \qquad (1)$$

, where $E$ is the energy of injected electrons in the measurement. All of the interparticle interactions that influence the electronic dispersion are summed up to yield the modified complex-valued self-energy $\Sigma$ of the (quasi)particles. In the quasiparticle approximation, the real part of the self-energy represents the change of the particle energy, and the imaginary part is proportional to the inverse of the particle lifetime (*1, 19*). The controlled observation of this modification in the electronic dispersion would be one of the most direct ways for quantifying and understanding the dynamics of an interacting many-particle system.

We extract the real and imaginary components of the self-energy $\Sigma$ ($Re\ \Sigma$ and $Im\ \Sigma$) by determining the peak locations and widths of momentum distribution curves (MDCs) of the spectra (see the description in **Fig. 2E-F**). In **Fig. 2F**, the extracted $Im\ \Sigma$s at various densities show step-like features at $\hbar\omega_{LO} \sim 36\ meV$ in the 1st subband and at $E = \varepsilon_{2,0} + \hbar\omega_{LO} \sim 70 - 80\ meV$ in the 2nd subband (see red and blue dashed lines). A step-like feature in the imaginary part of self-energy generally arises as a consequence of electrons interacting with a nearly dispersionless bosonic mode (*1*, *18*). This strongly suggests that these features in the self-energy are due to an electron-phonon interaction in GaAs.

To further corroborate the phonon effects, we have modeled a self-energy $\Sigma$ that includes impurity broadening and e-ph interactions in a weak-coupling limit (i.e. a perturbation theory) to calculate the simulated tunneling spectrum in **Fig. 2D** (see also **Fig. S4**). The simulation involves dispersionless longitudinal-optical (LO) phonons interacting with electrons via the Frohlich Hamiltonian (*20*), appropriate for a polar semiconductor such as GaAs (*1*). The simulated spectrum displays key aspects of the measured spectrum, and we can identify kinks and modifications (indicated by red and blue arrows) of the spectrum generated by the e-phonon coupling; the comparison indicates that the features located near 36 meV and 70~80 meV (depending on the density of the target QW) in the spectrum originate from the electron-phonon interaction combined with a specific subband structure.

Our data display the effects of LO phonons mixing different 2D subbands, which are otherwise independent in the single particle picture. The data also show these features appearing at various fixed energies, i.e. LO phonon energy (~36 meV) above subband bottoms as well as above the Fermi energy (see diagrams in **Fig. S4**). This is in contrast to prior ARPES measurements that have only detected effects from phonons referenced to the Fermi energy (*7*).

When the electron density of the target QW is further decreased to near depletion, electrons localize in minima of a disordered potential and can no longer screen interparticle interactions. Then, the validity of the perturbative calculations of self-energy due to the e-ph interaction also becomes questionable (see section 4 of Supplementary Materials). In this regime, the spectrum undergoes a qualitative change with the quasiparticle dispersion developing an unexpectedly sharp kink (circle in **Fig.2C**). The kink structure only exists at very low density near the depletion regime and low temperature, unlike other structures mentioned previously. From the shape of the kink we conjecture there exists a strong electron-boson coupling emergent near electron depletion (*21*). The energy of the kink is located at $\hbar\Omega \sim 48\ meV$ above the first subband minimum, where $\hbar\Omega$ is the energy of the conjectured bosonic mode. We note that AlAs-like phonons exist at ~49 meV in $Al_{0.8}Ga_{0.2}As$ with an e-ph coupling constant larger than that of GaAs (*22*). We also find this emergence of the kink coincides with an abrupt decrease in compressibility in a separate measurement (see **Fig. S2**). As localization arises in this regime, the onset of this feature may indicate the development of strongly-bound states composed of a localized electron and a phonon, i.e. polarons (*23*). This idea is supported by the observation that the kink structure disappears above $T \sim 16\ K$ (see **Fig. S8**), suggesting that the confinement from the disorder potential disappears by thermally activated electron hoppings. Note that study of the spectral function in this insulating regime is possible only with the capability of MERTS to measure unoccupied states in insulating systems.

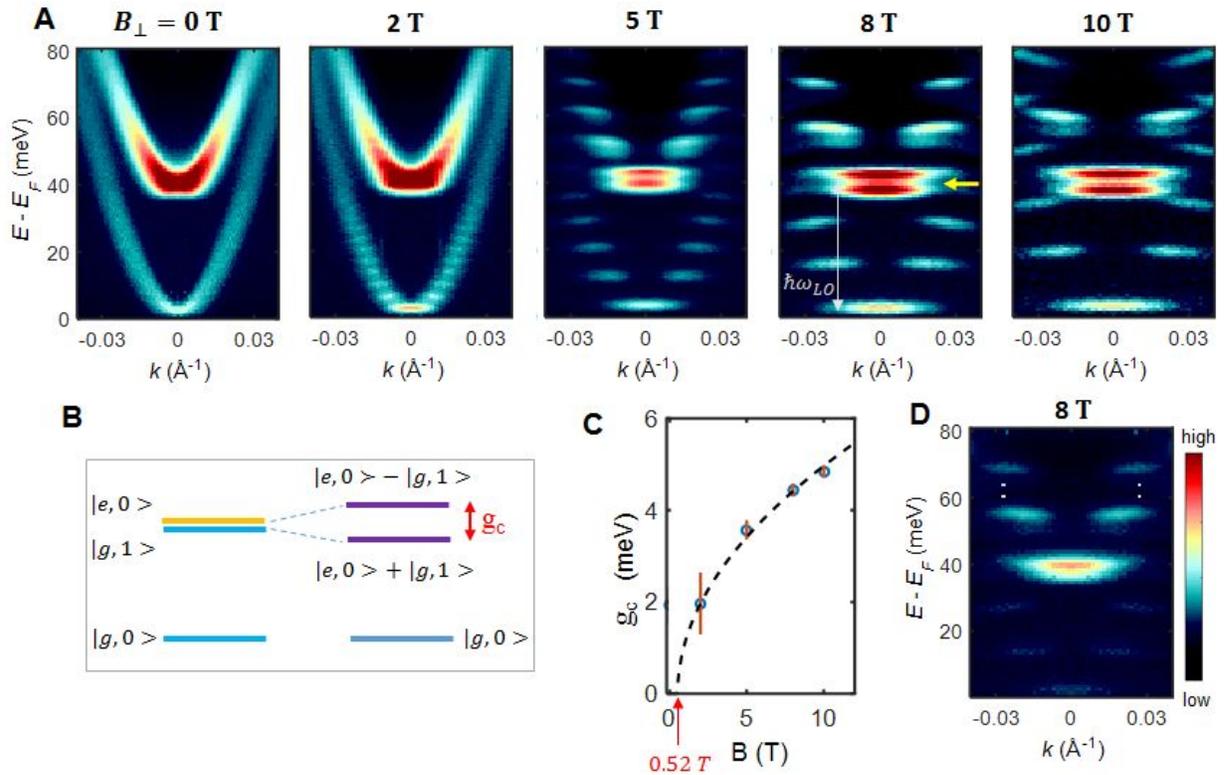

**Figure 3 | Evolution of E-k spectra under perpendicular magnetic fields.** **(A)** Spectra of the 2D electron system at a density of $n \sim 2.5 \times 10^{10}$ $cm^{-2}$ for various perpendicularly applied magnetic fields. With increasing field, Landau quantizations with the cyclotron energy spacing $\hbar\omega_c$ become clearly apparent. At $E - E_F$ ~ 40 meV, the spectrum displays a level splitting in the 5, 8, and 10T data. Tunneling into Landau levels produces features resembling line segments. The features resulting from tunneling into Landau index greater than zero appear as line segments at nonzero k that are slanted away from horizontal in the 5, 8, and 10T data. The slants arise from the non-zero width of our square well QWs (to be published). **(B)** A simple model that explains the level splitting. When the LO phonon energy matches the energy difference of 1st (blue lines) and 2nd subband (orange line), a resonance occurs (purple lines) to produce the strongly-coupled limit of an electron and a phonon, i.e. a polaron. The splitting arises because a long-lived composite state of the zeroth LL in the lowest (1st) subband together with a LO phonon (i.e. $|g,1>$) forms and resonantly interacts with the zeroth LL of the 2nd subband ( $|e,0>$). **(C)** The splitting energies (blue circles) and errorbars (vertical orange lines) are plotted as a function of magnetic field. The dashed line is a curve-fit of the data points at non-zero $B$ (from 2 T to 10 T) to $g_c = a \times \sqrt{B - B_0}$ with fixed $B_0 = 0.52$ $T$ (see main text). **(D)** The splitting diminishes greatly for a spectrum measured at higher density $n \sim 1.3 \times 10^{11}$ $cm^{-2}$ at 8T. Color scales in images in (A) and (D) are calibrated so that the same currents give rise to the same colors and intensities (see also Fig. S9).

To further investigate the effect of confinements that could lead many-particle interactions into the non-perturbative limit, we generated another means of confinement of electrons by applying perpendicular magnetic fields. As the magnetic field strength increases, the Landau quantization strongly modifies the dispersion into discrete Landau levels (LL) (**Fig. 3A**). Because electrons are confined in a length scale of the magnetic length $l_b$, the dispersion spreads over a finite extent in the momentum direction whose magnitude is proportional to $1/l_b$.

At $B_\perp > 0.52T$, all the electrons in the target QW fall into the lowest Landau level ($\nu < 2$). Near 36 meV ($\sim \hbar\omega_{LO}$) above the bottom of the unoccupied bands, the spectrum shows a surprisingly clear level splitting (see yellow arrow). We explain this as the polaronic effect dramatically enhanced to the point where the LL of the 2nd subband resonantly interacts with the LL of the 1st subband via a LO phonon. This is the signature behavior of the vacuum Rabi splitting - in this case, a "strong coupling" state between the zero-phonon and one-phonon states (*24, 25*) (see **Fig. 3B**) - and cannot be explained by any inelastic process involving simple phonon scatterings. Upon increasing the density of the 2D system so that only few empty states remain available in the ground LL, the splitting disappears (**Fig. 3D**; see also **Fig. S9**), consistent with the Rabi splitting picture. We realize an interesting analogy between the role of the electron density in polariton-cyclotron resonance experiments (*25, 26*) and that of the density of empty states $N_{empty}$ in the lowest LL in our measurement. This claim is quantitatively supported by the observation (**Fig. 3C**) of the splitting magnitude $g_c \propto \sqrt{N_{empty}} \propto \sqrt{B - B_0}$. Here, $B_0 = 0.52\ T$ corresponds to the field value at which electrons fully occupy the lowest LL, and roughly consistent with the density $n \sim 2.5 \times 10^{10}\ cm^{-2}$ determined by capacitance measurements. In this picture, the massive degeneracy created by the flattening of the electron energy dispersion and the spatial confinement due to a perpendicular magnetic field contribute to the formation of the strong coupling state. We find it remarkable that, as we tune perpendicular magnetic fields, the electronic spectral function shows the transition from the weak limit (the self-energy treatment at $B = 0$) to the strong limit of e-ph interaction. This observation constitutes the first direct measurement of the alteration of the electron energy spectrum in an equilibrium state due to the presence of ions in strong coupling. Such a measurement has been impossible in ARPES due to the requirement of application of high magnetic field. We anticipate even stronger polaronic effects in other semiconductor systems, such as hole-doped GaAs, CdTe, and $SrTiO_3$ that have stronger electron-phonon interactions.

Theory predicts that e-e interactions lead to changes in thermodynamic quantities (*1, 27*), but these effects are subtle and the measurements are difficult to interpret (*28*). MERTS provides a direct means for observing these interactions in high resolution spectra at near the Fermi energy, where LO phonons have small impact and the dominant modifications to the spectra arise from e-e interactions. In **Fig. 4**, we examine closely the spectra near the Fermi level. In **Fig. 4B**, at all densities, the measured $|Im\,\Sigma|$ reaches the minimum value near the Fermi level in accord with the phase space restriction argument of Fermi liquid theory: the inverse of the quasiparticle lifetime $1/\tau_q$ (or $2|Im\,\Sigma|$) due to e-e interaction decreases to zero quadratically in energy near the Fermi level (*30*). Due to disorder, however, the inverse lifetime has a non-zero value even at $E_F$ (*31*).

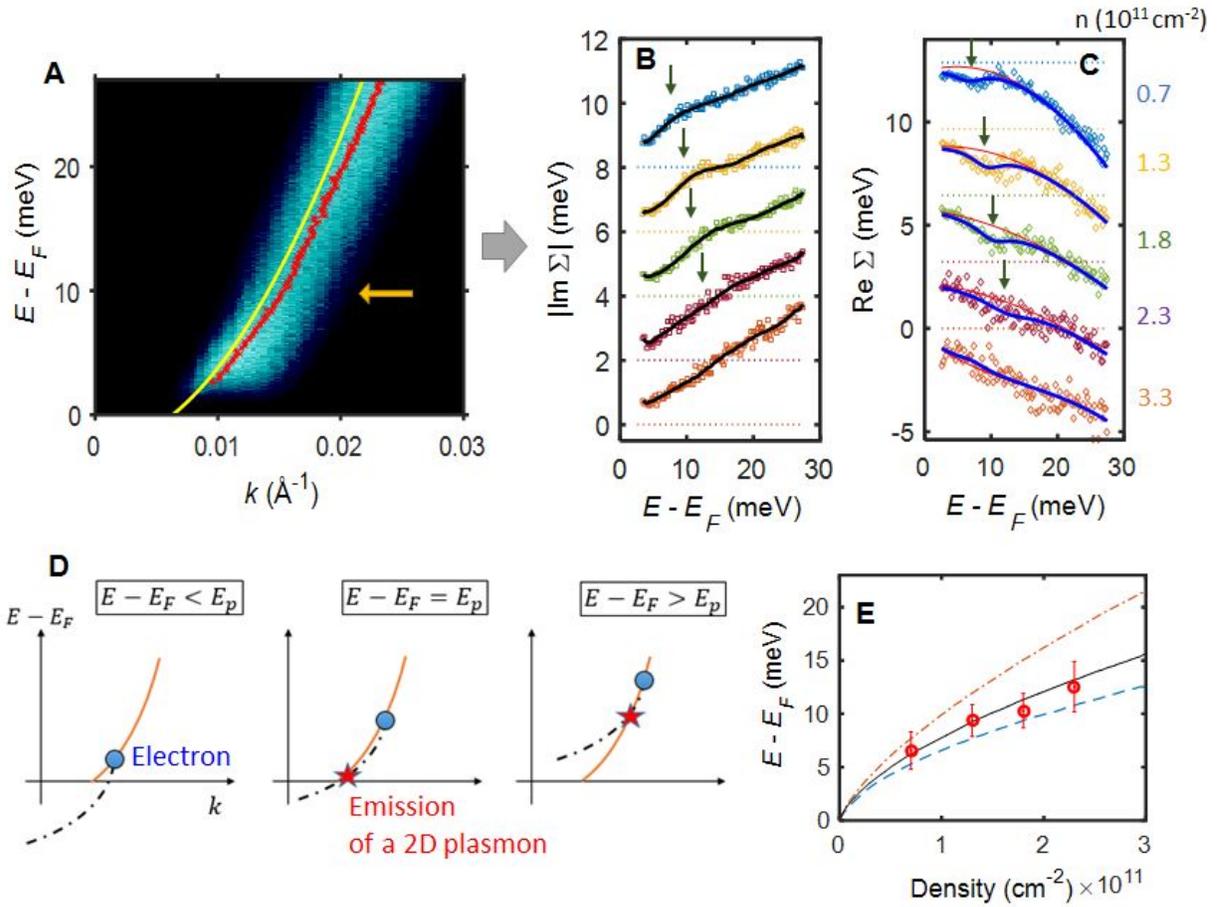

**Figure 4 | Electron-electron interaction and plasmon scattering. (A)** A zoom-in of the MERTS spectrum of the yellow rectangle of Fig.2E measured at 3K. The yellow curve shows the expected bare band dispersion, and red dots are measured peaks from fits to the data. There is a subtle, but statistically significant, kink in the red data points indicated by the arrow. **(B-C)** $Im\,\Sigma$ and $Re\,\Sigma$ from the spectra at various densities. In (C) the red curve represents a slowly varying background signal obtained by curve-fitting the data away from the kink structure to a 3rd order polynomial, and the blue curve is a curve-fit to the data with additional Lorentzian term in the curve-fit. The arrows in (C) point to the center of this Lorentzian. The same arrow locations are produced in (B), denoting a corresponding broad step feature in $Im\,\Sigma$. **(D)** Electron-plasmon interaction. The orange curves represent electron dispersion, and dash dotted curves visualize final electron states in event of emitting a plasmon. There exists a threshold energy $E_p$ above which an injected electron (blue dots) can lose energy by scattering a plasmon, and decay to lower energy states that exist above $E_F$ (red stars). **(E)** The locations of the features in (B-C) and how they change as a function of density. The red circles and error bars indicate peaks and widths of the Lorentzian function used to fit the data in (C). Also, plotted are theoretically expected curves following ref. (*29*). The blue dashed curve is from a semiclassical calculation, the red dot-dashed curve is from a random-phase-approximation (RPA), and the black solid curve is for a RPA that accounts for the non-zero well-width.

A closer inspection of the extracted self-energy at low energies, in **Fig. 4B-C**, shows additional structures that strongly depend on density (see vertical arrows) whose qualitative evolution as a function of density suggests an origin from e-e interactions. The step-like features in $Im\,\Sigma$ and dips in $Re\,\Sigma$ can be explained as electrons being scattered by plasmons. The strength and locations of such features that vary as a function of electron density $n$ is in quantitative agreement with the theory of plasmons in 2D (*5, 6, 29*). In this picture, the density dependence comes from a combination of a varying strength of the Coulomb screening by neighboring mobile electrons and a change of plasmon dispersion itself (see also **Fig. S6**). There has been no previous report of measuring the plasmonic effect in the spectral function in a system of massive 2D electrons. The plasmons modify the electron dispersion near the Fermi level and show the square-root density dependence intrinsic to 2D plasmons (**Fig. 4E**). Our data thus present a direct measurement of quantities that were only theoretically investigated previously (*5, 6*), demonstrating that the e-e scattering is suppressed near the Fermi level until plasmons significantly contribute scattering processes. We note that the effect of plasmon scatterings have been observed in the electron spectral function of massless electrons in graphene using ARPES (*32*), but the structure did not directly reflect the plasmon dispersion relation. Our ability to determine the plasmon dispersion in the low density regime can impact the understanding of topics of significant recent interest. For instance, Ruhman and Lee proposed (*33*) that the electron-plasmon coupling in a low density regime in $SrTiO_3$ could act as the pairing glue of the superconductivity.

We have developed a new method (MERTS) that for the first time allows a high resolution and full determination of spectral function of a high-quality semiconductor 2D electronic system. The results display the effects of interparticle interactions that dramatically modify the electronic *E-k* dispersion spectrum as a function of density and magnetic fields. Under high magnetic fields, the technique can allow visualization of anisotropic wave functions of strongly correlated phases in the quantum Hall regime and providing crystallography of stripe, bubble phases and Wigner crystals (*34*). Importantly, MERTS is compatible with many embedded structures including electronic devices and circuits, and highest quality heterostructures for fundamental research, thus opening a new way of investigating physics of interacting particles in previously challenging geometries (*35*).

**Acknowledgements**

The work at MIT was funded by the BES Program of the Office of Science of the US DOE, contract no. FG02-08ER46514, and the Gordon and Betty Moore Foundation, through grant GBMF2931. The work at Princeton University was funded by the Gordon and Betty Moore Foundation through the EPiQS initiative Grant GBMF4420, and by the National Science Foundation MRSEC Grant DMR-1420541.

# Supplementary Materials

## 1. *Materials and Methods*

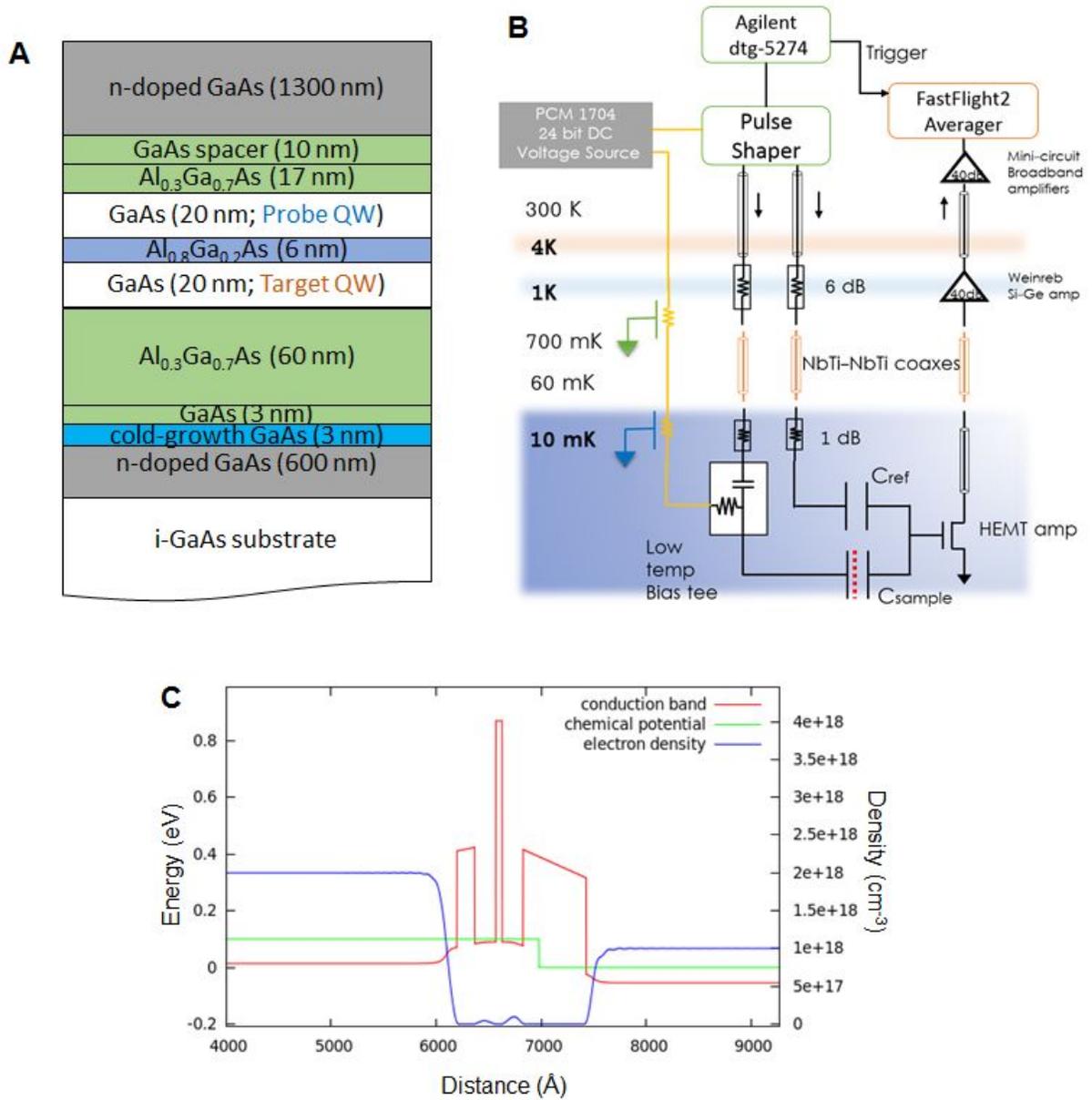

**Figure S1 | Experimental details.** **(A)** The wafer growth sheet. **(B)** A schematic of the experimental setup (*36*). **(C)** A numerical simulation of the electron density and electric field distribution of a 2D-2D QW structure based on a self-consistent solution of the Poisson-Schrodinger equation under external DC voltage.

We use photolithography and etching to define our sample a mesa (~ 150 μ*m* diameter) froma wafer of an MBE grown GaAs/AlGaAs heterostructure with profile shown in Fig. S1A. The top and bottom thick layers (>500 nm) form 3D electrodes and are doped with Si with the density of approximately $2 \times 10^{18}$ *cm*$^{-3}$. We perform numerical simulations of the electron density distribution along z-axis (the growth direction) of the structure by self-consistently solving the 1D Poisson-Schrodinger equation, leading to results such as those shown in Fig. S1C. The thickness of the GaAs spacer layer sets the density of the probe QW. In the present measurements, the thickness of the spacer is 10 nm, and the electron density induced into the probe QW is $n \sim 3 \times 10^{10}$ *cm*$^{-2}$. The simulation shows that, in applying a DC bias voltage $V_b$ across the electrodes the density of the target QW changes, but the density of the probe QW effectively remains same until the target QW nearly fully depletes. We determine the density of target QW vs. bias voltage by integrating the measured compressibility (i.e. capacitance) of the QW (Fig. S2A).

We perform cryogenic broadband time-resolved measurements with a sample mounted in a rotational stage at the mixing chamber of a $^3$He-$^4$He dilution refrigerator, and apply magnetic fields using a 16T superconducting magnet. To thermally protect the sample from the warm outside environment and preserve the low electron temperature, we use a combination of stainless steel, copper and NbTi superconducting coaxes with in-line attenuators (see Fig. S1B). For measurements of tunneling spectra, we apply repetitive square pulses to induce the tunneling voltage between two QWs. An initial voltage pulse whose amplitude is $V_p$ defines the actual energy of tunneling electrons. The energy is determined by accounting for the geometric lever-arm of the heterostructure - well-to-well distance divided by the distance between electrodes - e.g. $E_t = eV_t \sim (26\ nm/113\ nm) \times eV_p$. The initial pulse has a short duration of ~ 100 *nsec* to ensure that the tunneling current is measured before the QW densities significantly change. The initial pulse is followed by a voltage pulse with an opposite polarity, which retrieves the tunneled electrons back to the probe QW. After these pulses, we wait for ~ 300 μ*sec* to further ensure that the whole system is back to equilibrium. As shown in Fig. S1B, we use HEMT amplifiers to detect the tunneling charges capacitively. The signal from the HEMTs is amplified by a SiGe cryogenic low-noise amplifier (Weinreb amplifier) at the 1K pot, and finally recorded by a high speed signal averaging oscilloscope (SignalRecovery FastFlight2). We average an order of ~10,000 pulse sequences to obtain individual pixels in a 2D E-k spectrum. We scan the energy axis of the spectrum by sweeping the pulse amplitudes using a DC voltage source multiplexed with a digital timing generator (Tektronix DTG-5274). The scan along the momentum axis is controlled by stepping the predetermined combination of magnetic fields and a rotation angle of the sample stage. A complete spectrum of a single bias, such as shown in the main text, usually takes ~15 hours to acquire with a computer automated system.

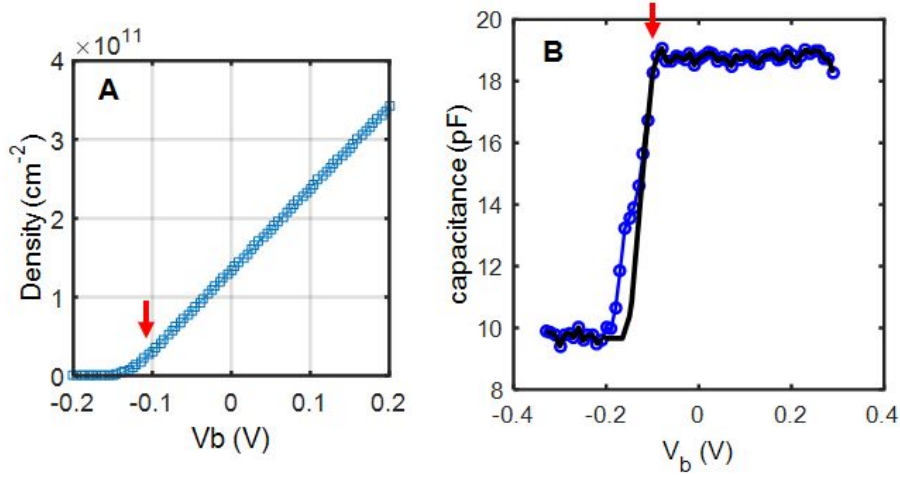

**Figure S2 | Density determination using capacitance measurement.** The red arrows point to the bias voltage value ($V_b \sim -0.11\ V$) below which the electron system transitions to the incompressible state, signaling particle localization. **(A)** The density of target QW vs. bias voltage estimated by integrating the measured compressibility (i.e. capacitance) of the QW, with calibration determined through measurements showing Landau levels in magnetic field. The density designated by the red arrow corresponds to $\sim 2.5 \times 10^{10}\ cm^{-2}$. **(B)** A representative total capacitance plot (blue circles) of two QW structure measured for density calibration. The black line is the estimated capacitance of the target QW, based on the total capacitance data. The difference between the blue and black lines is due to the screening effect of the probe QW in the capacitance measurement.

## 2. Tunneling conductance and momentum conservation

When two electronic systems are separated by a planar tunneling junction, the tunneling current is given by the following form,

$$I_t(V_t) = \Sigma_{kk'}|T_{kk'}|^2 \int_{-\infty}^{\infty} dE \int_{-\infty}^{\infty} dE'\, A_L(E,k)\, A_R(E',k') \times$$
$$[f_L(E,k_B T)\{1-f_R(E',k_B T)\} - \{1-f_L(E,k_B T)\}f_R(E',k_B T)]\, \delta(E-E'-eV_t), \quad (S1)$$

where $A_L$ and $A_R$ are the spectral functions of left and right systems, respectively, with the following formula

$$A(E,k) = \frac{1}{\pi} \frac{|Im\, \Sigma|}{(E-\varepsilon_k - Re\, \Sigma)^2 + (Im\, \Sigma)^2}. \quad (S2)$$

And $f_{L(R)}$ is the Fermi distribution functions of the left (right) system, $\varepsilon_k$ is the bare electron band energy, and $\Sigma$ is the electronic self-energy. The $k$ represents planar momentum in the 2D plane perpendicular to the tunneling direction. With translational symmetry in the plane, the planar momenta are conserved in the tunneling process.

Under in-plane magnetic fields, the planar momenta are still conserved but with an offset given by $\Delta k_{x,y} = eB_\parallel d_t/\hbar$, as shown in Fig. S3. The measured tunneling spectrum resembles very closely the simulation using this simple picture, demonstrating the energy and momentum conservation in this experiment.

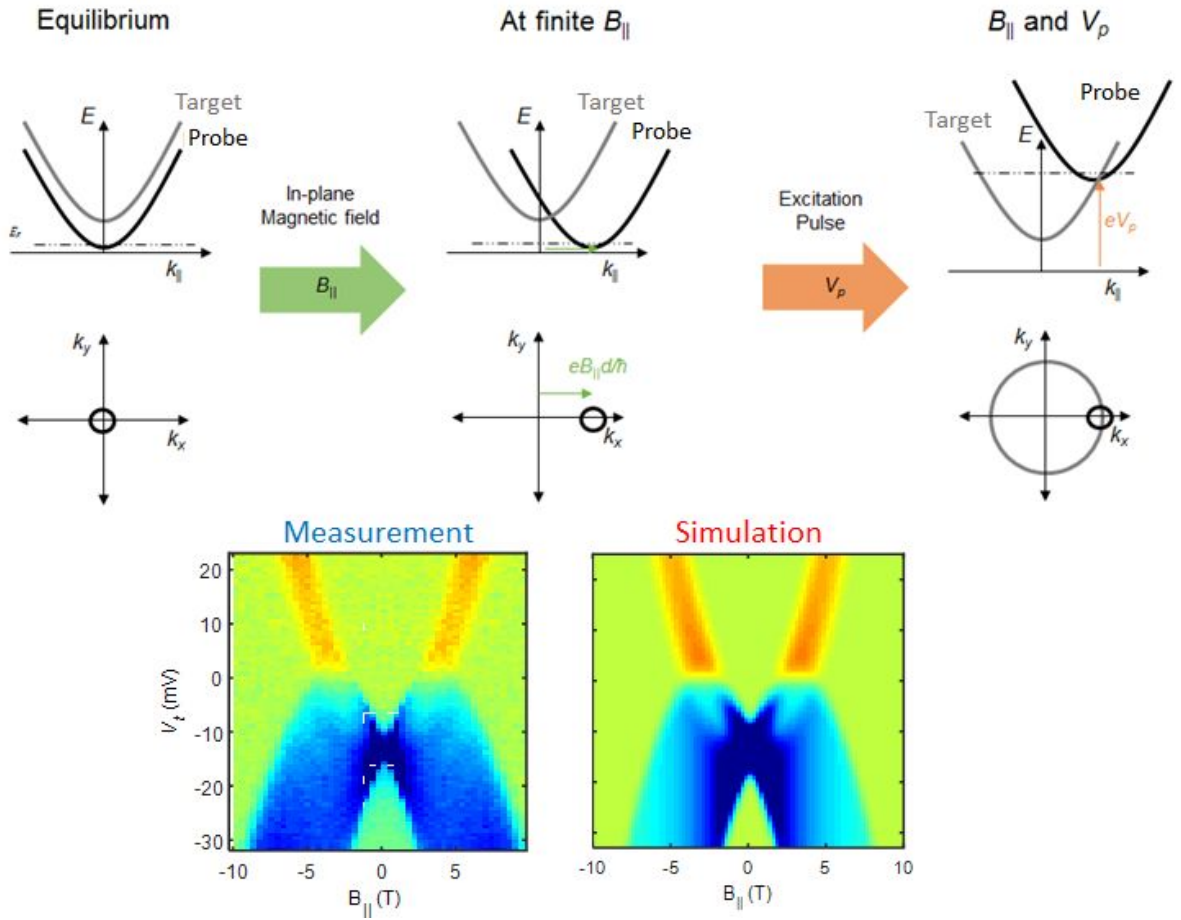

**Figure S3 | Energy and momentum selective tunneling processes confirmed by comparing measurement with simulation results.**

## 3. Many-particle interactions and self-energy

When particles in a system interact with each other, simple product states of individual particles are typically no longer eigenstates of the system. Instead, the interacting many-body system describes the particle-like excitations (quasiparticles) with finite lifetime and modified energy different from those of bare electrons (*1*, *2*). This picture is formally expressed in the spectral function of Eq. (S2), and the *self-energy* function $\Sigma$ has the essential information of the inter-particle interactions. Thus, measurements of the spectral function and the self-energy is of prime interest both experimentally and theoretically.

In the quasiparticle approximation (*7*, *17*), the essential information of the spectral function can be represented by two quantities: a quasiparticle lifetime and a modified energy. The lifetime of an electronic quasiparticle can be obtained from the inverse of the energy broadening $\Delta E$ of the energy distribution curves (EDC; curves cut at $K$ along the energy axis) of the dispersion, approximately an inverse of the imaginary part of the electronic self-energy. On the other hand, the renormalized energy of the quasiparticle comes from the locations of the peaks $E_{peak}$ of the quasiparticle distribution. If the self-energy does not change very rapidly in energy, the self-energy is obtained by solving these equations,

$$-2Im\,\Sigma = \Delta E \quad \text{and} \quad E_{peak} - \varepsilon_K - Re\,\Sigma = 0\,. \tag{S3}$$

## 4. Numerical simulation of tunneling spectrum

Before starting the procedure of extracting the self-energy, we can compare the measured spectral function to a simple theoretical model for a basic sanity check. With a simple model of the electron-boson coupling with no momentum dependence, an isotropic matrix element, and a small electron-boson coupling constant $\alpha$, the self-energy can be written in the limit of zero temperature as (*1, 16, 17, 37, 38*)

$$\Sigma(E_t) = 1/\pi \int_0^\infty dE'\, \alpha^2 F(E') \int_0^\infty dE''\, N(E'')/N_0 \left[\frac{\theta(E'')}{E_t + E' - E'' + i0^+} + \frac{\theta(-E'')}{E_t - E' - E'' + i0^+}\right], \tag{S4}$$

where $\alpha^2 F(E)$ is the Eliashberg function (*17*), $N(E)$ is the electron density of states, and $N_0$ is a the electron density of states of the non-interacting system at the Fermi level. Further assuming a single Einstein bosonic mode at $\hbar\omega_{ph}$, the self-energy can be easily calculated. The perturbation theory arguments leading to Eq. (S4) are generally considered valid when $\lambda \times \hbar\omega_{ph}/E_F < 1$ (*38*) where $\lambda$ is the Frölich dimensionless e-ph coupling constant (*20*). The value of $\lambda$ is ~0.07 in GaAs (*39*), and the above condition satisfies when electron density is not too low (i.e. $E_F >\sim 3\,meV$). Qualitatively speaking, these e-ph processes effectively change the electron dispersion to result in a lower group velocity near the Fermi level and thus make the effective mass of electrons heavier by the mass enhancement factor $\lambda = -(\partial Re\,\Sigma/\partial \varepsilon)^{-1}$ (*1*), which, in principle, one can calculate when the $Re\,\Sigma$ is accurately measured.

We combine equations S1, S2, and S4 to numerically simulate tunneling spectra as in Fig. S4A-B. We assume the 2D density of states of electrons with three subbands, and an Eliashberg function $\alpha^2 F$ of LO phonon. To produce a simulation that resembles our data, we take the optic phonon density of states to be centered at $\hbar\omega_{ph} = 36\,meV$ and $\alpha^2 F$ to be a Lorentzian with height $A = 3.5\,meV$ and width $W = 2.5\,meV$. We, then, calculate $Im\,\Sigma$ from Eq. (S4) using the identity $\frac{1}{E-E'+i0} = \hat{P}(\frac{1}{E-E'}) - i\pi\delta(E-E')$, where $\hat{P}$ denotes the principal part. This results in the curve for $Im\,\Sigma$ shown in Fig. S4B. The causality of the self-energy function in Eq. (S4) implies that $Im\,\Sigma$ and $Re\,\Sigma$ are related through the Kramers-Kronig relationship (KKR), thus we can evaluate $Re\,\Sigma$ from the $Im\,\Sigma$ as also shown in Fig. S4B.

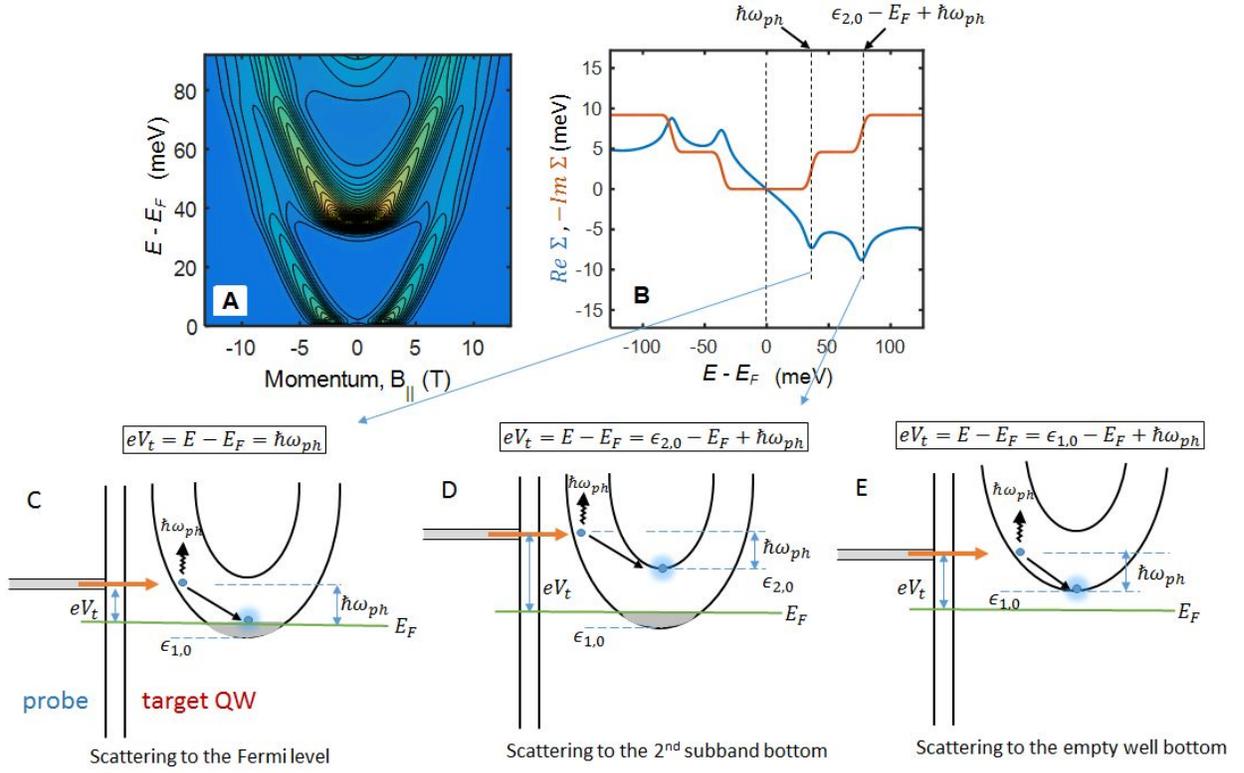

**Figure S4 | An example of tunneling current simulation with a simplified self-energy model included.** (**A**) Simulated tunneling spectrum. (**B**) The real and imaginary parts of a simulated self-energy used to simulate the spectrum in **A**. The real part is obtained from the imaginary part by the Kramers-Kronig relationship. Here, $\varepsilon_{2,0}$ is the 2nd subband minimum energies. (**C-E**) Schematics of various conditions of (virtual) phonon emissions that can occur and affect the electron spectrum and the self-energy. We can understand the outstanding features of electron-phonon interaction in the self-energy with the simplified mechanisms.

## 5. Extraction of self-energy from momentum distribution curves (MDCs)

In the current study, we use MDCs to extract the self-energy. There are a number of benefits from extracting the self-energy from MDC instead of EDC (*15*): (i) When the self-energy is mostly a function of energy, a subtle change of the self-energy is better resolved in MDC because each curve has a single energy value; (ii) near the Fermi level, EDC is strongly distorted due to the Fermi distribution, making an interpretation very complicated; and (iii) a potential energy dependent tunneling matrix term can heavily affect EDCs.

We assign $\Delta k$ to be the half-width of the MDC, and $k_{peak}$ to be the peak location of the MDC. Note that the MDC is measured at a certain energy value $E$. From Eq. (S3), we can easily show that

$$|Im\Sigma| \approx \Delta k \times \hbar v_{g0} \text{ , and } Re\,\Sigma = E - \varepsilon_{k_{peak}}.$$

Thus, we can extract the self-energy by determining the peak locations $k_{peak}$ and widths of the quasiparticle peaks $\Delta k$ of the MDCs from the measured spectrum.

In experimentally determining the $k_{peak}$ and $\Delta k$, the small FS of the probe layer is the key factor that makes the entire procedure simple and transparent. We can find the locations of $k_{peak}$ from measured MDCs very accurately over the entire range of experimental parameter space. In contrast, obtaining $\Delta k$ from the measured MDC widths requires more attention. The small but finite extent of FS of the probe layer still can generate a convolution effect. For example, in the limit where the intrinsic width of the MDC is much smaller than the momentum width of the FS of the probe layer, the convolution effect can be significant, and we need a method for deconvolution for a meaningful quantitative analysis. While the fully deconvolving the Eq. (1) is difficult and mathematically unstable, due to the small size of the FS of the probe QW we can use a good approximation to find a simplified version of the deconvolution, of which we later confirm the validity by checking with the KKR.

We expect the measured MDC width (the value determined by curve-fitting the measured tunneling spectrum in the horizontal direction in our plots) is always larger that intrinsic MDC width (the value solely given by the spectral function of quasiparticles) due to the finite extent in momentum of the FS of the probe layer. Note that the finite extent of the FS in energy (~0.5 meV) has negligible effect compared to the finite extent in momentum. As such,

$$\Delta k_{measured} = \Delta k_{intrinsic} + \delta.$$

By using the simulation described in the earlier section, we confirmed that $\delta$ is nearly constant throughout the energy range of the measurement, and the value is determined by the finite extent of the probe QW's FS. Then, the extraction of $Im\,\Sigma$ from the measured MDC width (HWHM) is a straightforward procedure as following

$$|Im\,\Sigma| = 1/Z_k \times (\Delta k_{measured} - \delta) \times \hbar v_g$$
$$\approx (\Delta k_{measured} - \delta) \times \hbar v_{g0}$$

, where $v_g$ is the group-velocity of the quasiparticle dispersion, $v_{g0}$ is the group-velocity of the bare electron dispersion, and $Z_k$ is the quasiparticle strength (*7, 17*). Note that the formula has the quasiparticle strength $Z_k$ explicitly written as we intend to demonstrate that the use of the bare group velocity is a better approximation in the formal treatment of the quasiparticle picture. We use $\delta \approx 0.0032 Å^{-1}$ estimated from the simulation based on the size of FS of the probe QW. In extracting $Im\,\Sigma$, we expect that these approximation procedures to give quantitatively good values within 10~20 % based on the simulation. On the other hand, the extraction of $Re\,\Sigma$, measured by locating peaks of the measured MDCs, is good within a few percent or less given that the bare band dispersion is known.

In extracting the $Re\,\Sigma$, we used $m_{eff} = 0.067\,m_e$, and non-parabolicity parameter $\eta \sim -43.7\,Å^2$ with $\varepsilon_k = \hbar^2/2m_{eff} \times |k|^2(1 + \eta|k|^2)$ to calculate the expected bare electron dispersion $\varepsilon_k$. The tunneling distance $d_t$ is a source of possible systematic error in this estimation because the $k$ is calibrated with $d_t$. The $d_t$ is determined from the 1D Poisson-Schrodinger equation to account for the well-to-well distance change when a voltage bias is applied across the top and bottom electrodes. Generally, the $d_t$ changes less than 1% when the tunneling energy $E_t = eV_t$ changes, and less than 5% when density changes upon application of DC voltage bias. It is important to account for this effect when we extract $Re\,\Sigma$, otherwise the effect generates an erroneous background slope in $Re\,\Sigma$.

To confirm the validity of the extraction of self-energy, it is important to check the extracted $Im\,\Sigma$ and $Re\,\Sigma$ are consistent with each other; the imaginary and real part of the self-energy are mutually related via the Kramers-Kronig relationship (KKR), a formal requirement for causality of response functions of a physical system. As shown in Fig. S5C, we plotted $Re\,\Sigma$ converted by the KKR from the $Im\,\Sigma$ of the measured MDC widths (black curves), and colored data points that represent the values measured by the energy difference of measured peak locations to the bare dispersion, i.e. $E_{MDCpeak} - \varepsilon_k$. These two curves show a reasonable agreement with each other. Thus, we conclude that values extracted from our tunneling spectrum validly represent the intrinsic self-energy of an interacting electronic system.

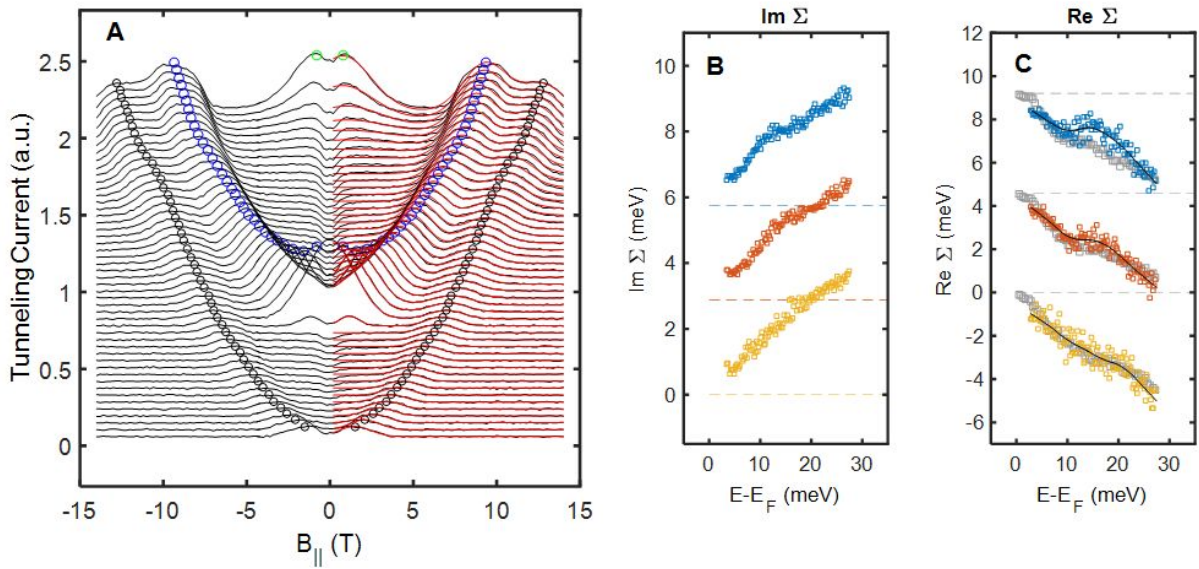

**Figure S5 | Extraction of the self-energy from momentum distribution curves.** **(A)** MDCs of the measured tunneling spectra (black lines) and curve fits (red lines) for extracting peak locations and widths. **(B)** $Im\,\Sigma$ obtained after the simple deconvolution process discussed in the section 5. The curves are offset for clarity. Carrier densities are 1.2 (blue), 1.7 (red), and $2.2 \times 10^{11}$ $cm^{-2}$ (yellow) from top to bottom curves. **(C)** A comparison between the $Re\,\Sigma$ obtained by using the KKR on the data of (B) (colored markers) and the $Re\,\Sigma$ obtained by directly measuring $E_{peak} - \varepsilon_k$ in the spectrum (gray markers). Black lines are moving averages of the colored data points. All curves are offsetted for clarity with individual zero values indicated by horizontal dashed lines.

## 6. Neutral excitations of a 2D system upon a sudden injection of an electron

Many phenomena inside of a solid originate from the many-particle interactions of electrons, ions and their collective excitations. One of the most intuitive ways to describe the interactions is in terms of a scattering process between two particles that increases the imaginary part of the self-energy (most intuitively as an increase in the inverse quasiparticle lifetime). By enumerating possible scattering processes, we thus can predict basic features of the self-energy to gain insight of the underlying interactions.

Electrons injected into a 2D electronic system at energy $eV_t = E_t = E - E_F$ decay towards the Fermi level, generating various excitations of the subject 2D system by means of scattering events. One example of visualizing the phase space of possible scattering events is shown in Fig. S6. Because in the scattering process the momentum and energy are always conserved, the scattering is only possible when the momentum and energy of excitations (Fig. S6A) match the momentum and energy of energy loss mechanisms (Fig. S6B). The dispersion of the electron system, the injection energy, and the Fermi level define the phase space for possible inter-particle interactions (Fig. S6C).

As the injection energy of an electron increases, in general, more scattering channels open and the quasiparticle lifetime decreases ($2|Im\ \Sigma|$ increases) assuming that the scattering matrix element roughly stays constant as a function of energy and momentum. Thus, the imaginary part of the self-energy, inversely proportional to the lifetime, shows jumps and increasing slopes that can be identified with the specific scattering mechanisms (Fig. S6C). Noticeably, electrons strongly scatter plasmon and phonon at characteristic energies (black dash-dot lines), but also generate e-h continuum excitations in a wide range of energy (blue-green shaded regions). In $Im\ \Sigma$, scatterings with particles with well-defined energy, such as plasmons and phonons, show up as steps, and the generations of e-h continuum excitations usually show up as increasing slopes (Fig. S6D).

In a QW, there are multiple subbands exist due to the non-zero width of confinement. This leads to two different kinds of e-h continua (*40*); intrasubband (blue shaped region) and intersubband (green shaded region) e-h continua. The former affects electrons near the Fermi level, but is known to be suppressed near the Fermi level due to the Pauli blocking that restricts phase space for e-e scatterings. The intersubband e-h continuum excitations, on the other hand, contribute to the electron quasiparticle spectrum above $eV_t \sim \varepsilon_{2,0} - E_F$ for decays that leave one electron into the second subband. Above another threshold at $eV_t \sim 2(\varepsilon_{2,0} - E_F)$, the decay of one high energy electron in the first subband can place two electrons in the second subband.

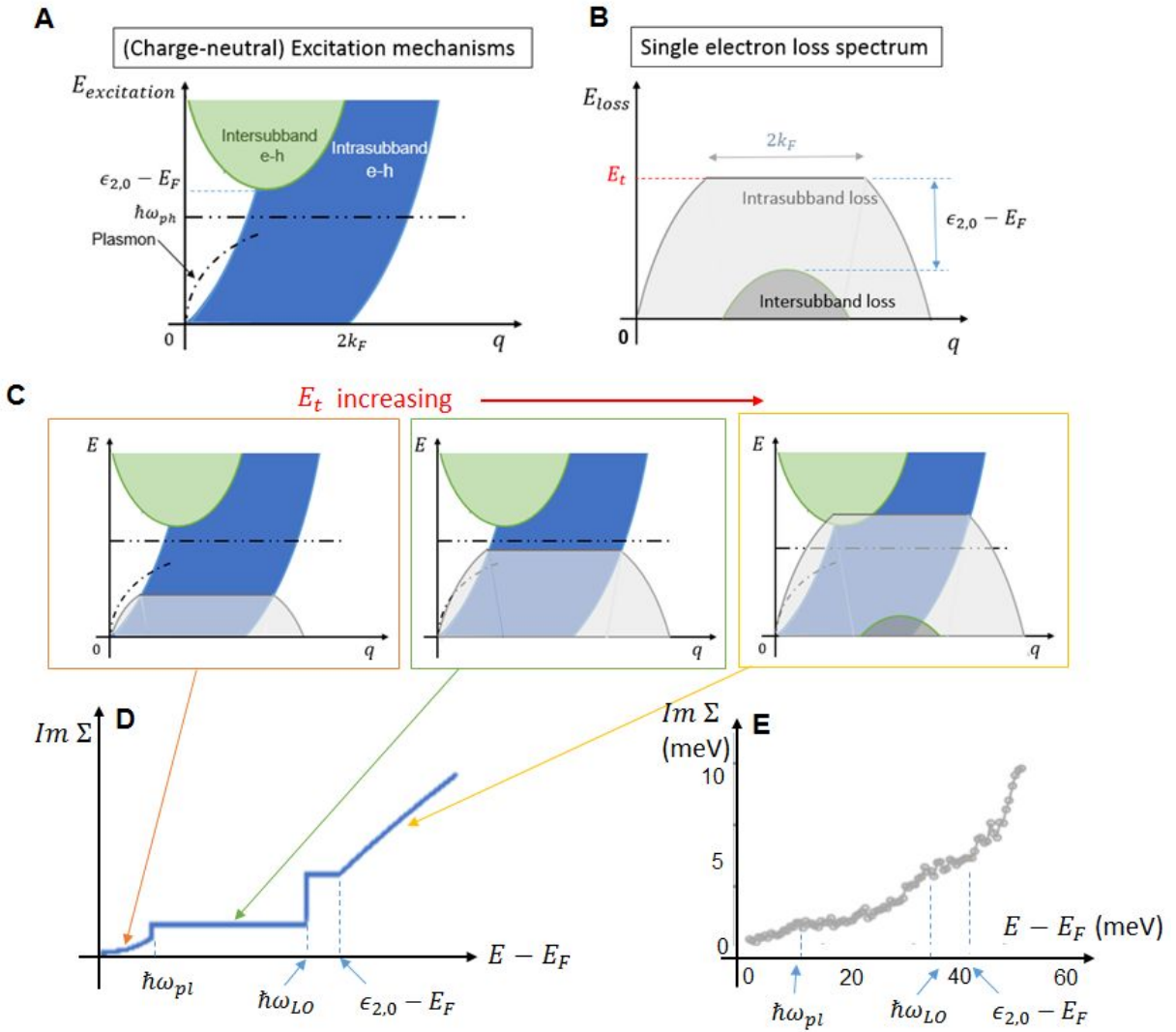

**Figure S6 | Possible excitation mechanisms and loss channels of a 2D electron system.** **(A)** Green and blue shaded regions indicates (E,q) values of possible intra- and intersubband e-h excitations. Black dash-double-dot and dash-dot lines show (E,q) values of phonon and plasmon excitations, respectively, as denoted in the figures. **(B)** Light and dark gray shaded regions indicate possible (E,q) values that an injected electron can lose energy and momentum (i.e. electron-loss spectrum). **(C)** When the electron-loss spectrum overlaps in E-k space with the excitations spectra, there is a non-zero probability of an injected electron at energy $E_t$ to lose energy and momentum by generating excitations as listed in (A). **(D)** Qualitative prediction of $Im\,\Sigma$ of the quasiparticle dispersion in the 1st subband based on the above model. **(E)** Comparison of the prediction with the typical data measured at $n = 0.7 \times 10^{11}\ cm^{-2}$. Note the qualitative agreement of features in the prediction (D) and the measurement (E).

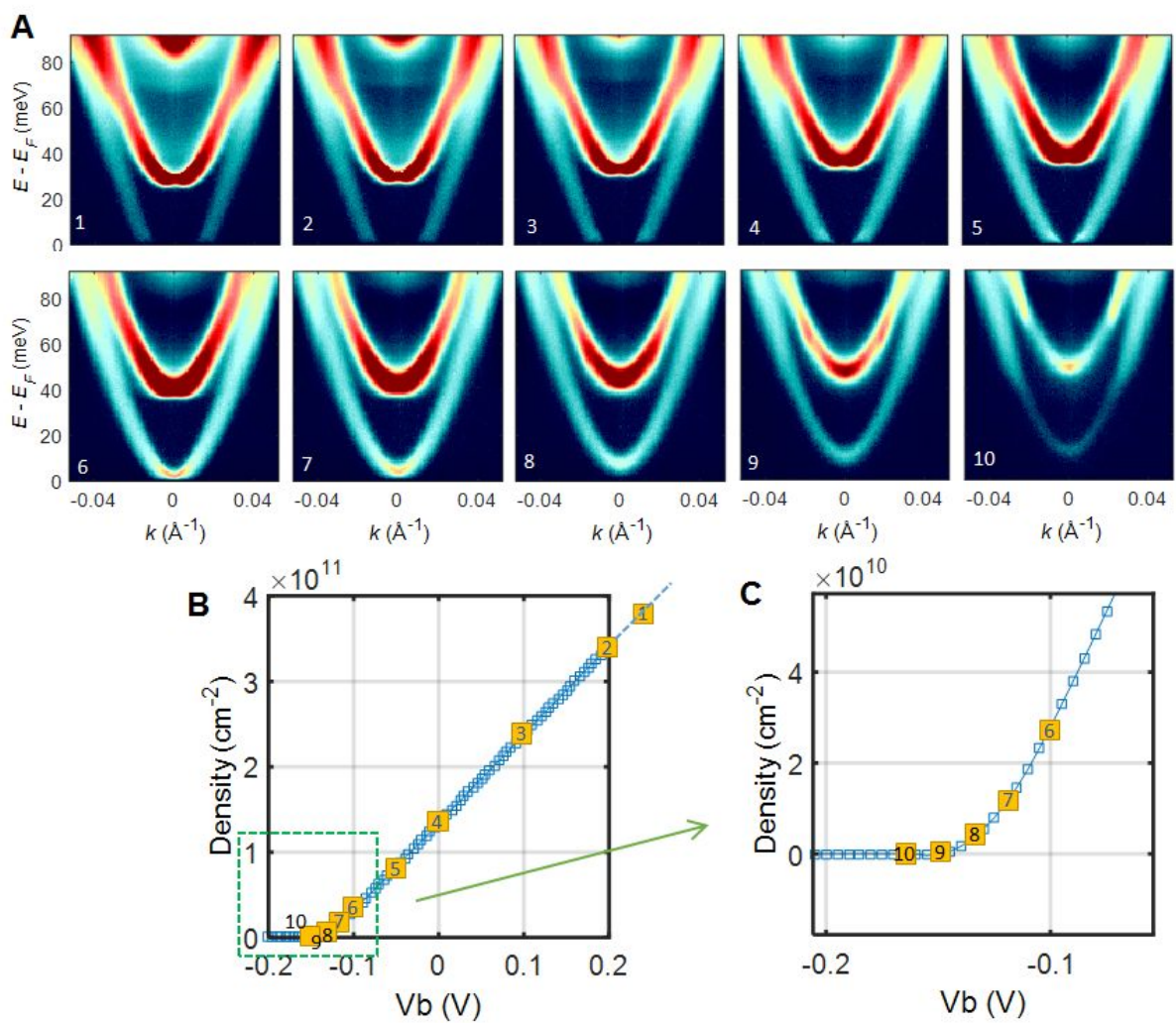

**Figure S7 | Measured spectra at T = 1.5 K at various electron densities. (A)** We show measured tunneling spectra at various densities, whose values are indicated by the corresponding numbered yellow squares in the density vs. bias voltage plots in **(B)**. **(C)** The plot is a zoomed-in view of the green rectangle in (B).

(continued on next page)

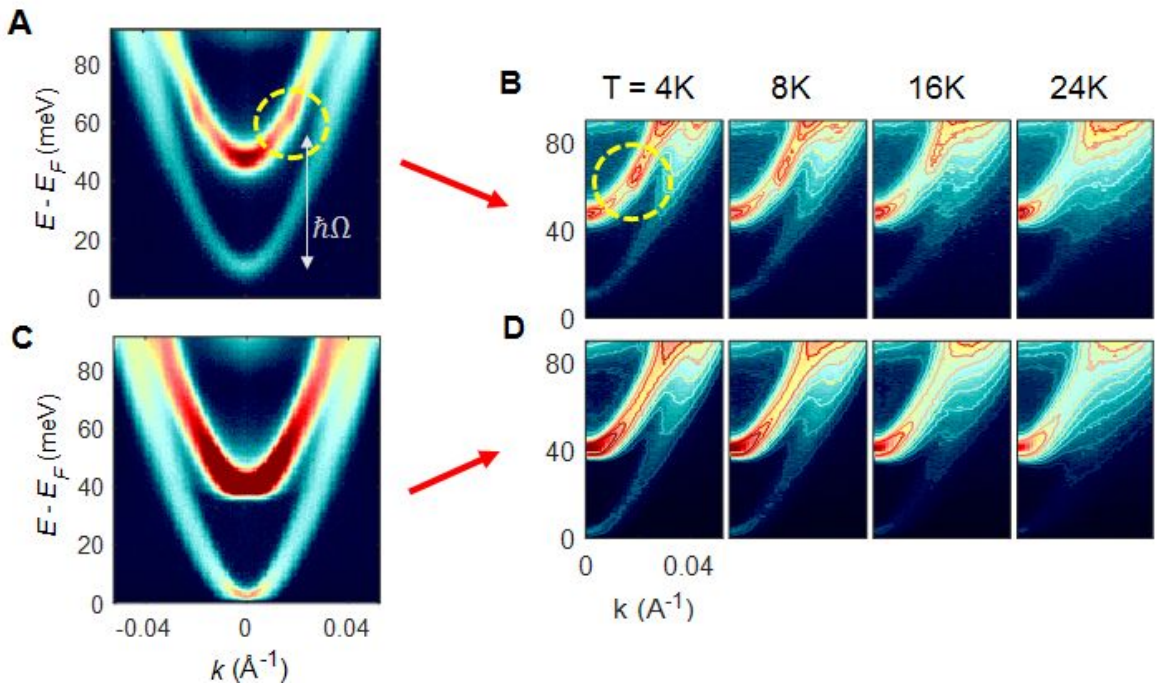

**Figure S8 | Temperature dependence of the kink structure.** **(A)** A tunneling spectrum measured at nearly zero density and $T = 1.5\ K$, and **(B)** spectra measured at 4, 8, 16 and 24 K with the same density. **(C)** A spectrum measured at $n \sim 3 \times 10^{10}\ cm^{-2}$ and $T = 1.5\ K$, and **(D)** spectra measured at 4, 8, 16 and 24K at the same density. The kink structure (inside the yellow circle) only exists at very low density near the depletion regime and low temperature below 16 K. In (B) and (D), we adjusted color contrast and added constant current contours to ease distinction of the kink features as temperature varies.

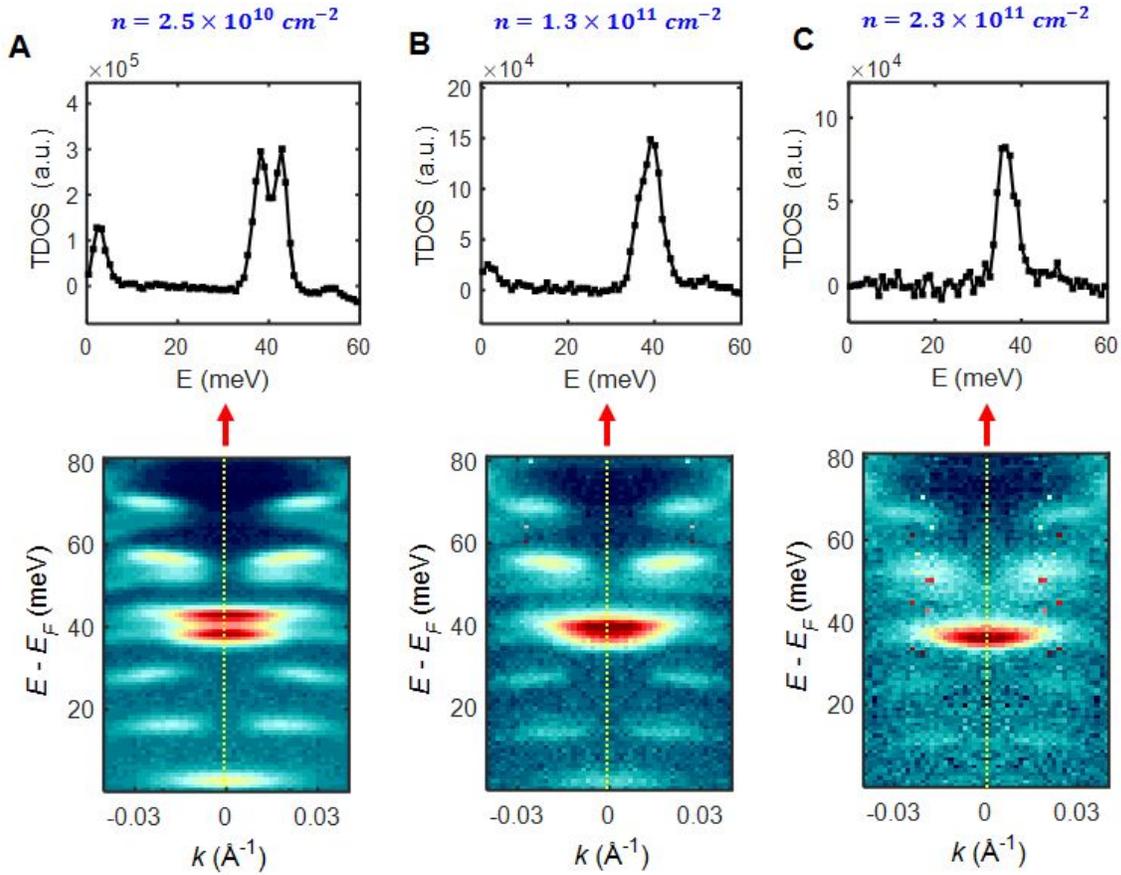

**Figure S9 | Line-cuts of the resonant polaron splitting at $B_\perp = 8T$. (A)** $n = 2.5 \times 10^{10}\ cm^{-2}$ **(B)** $n = 1.3 \times 10^{11}\ cm^{-2}$. **(C)** $n = 2.3 \times 10^{11}\ cm^{-2}$. Each upper panel shows a vertical line-cut through the yellow dotted lines in each spectrum. The splitting disappears at higher electron densities because the unoccupied electronic states near the Fermi level decrease and eventually no more 1-phonon states are available for the polaron resonance to occur. The sample temperature is 3K. In each 2D spectrum, the contrast is adjusted for maximum visibility.